# Quantum oscillation evidence of topological semimetal phase in ZrSnTe

Jin Hu[1,2*], Yanglin Zhu[2], David Graf[3], Zhijie Tang[2], and Zhiqiang Mao[2†]

[1] Department of Physics, University of Arkansas, Fayetteville, Arkansas, 72701

[2] Department of Physics and Engineering Physics, Tulane University, New Orleans, Louisiana, 70118

[3] National High Magnetic Field Laboratory, Tallahassee, Florida, 32310

Abstract

The layered *WHM* - type (*W*=Zr/Hf/La, *H*=Si/Ge/Sn/Sb, *M*=S/Se/Te) materials represent a large family of topological semimetals, which provides an excellent platform to study the evolution of topological semimetal state with the fine tuning of spin-orbit coupling and structural dimensionality for various combinations of *W*, *H* and *M* elements. In this work, through high field de Haas–van Alphen (dHvA) quantum oscillation studies, we have found evidence for the predicted topological non-trivial bands in ZrSnTe. Furthermore, from the angular dependence of quantum oscillation frequency, we have revealed the three-dimensional Fermi surface topologies of this layered material owing to strong interlayer coupling.

*jinhu@uark.edu

†zmao@tulane.edu

Three-dimensional (3D) topological Dirac and Weyl semimetals possess symmetry protected linear crossings near the Fermi level, harboring relativistic Dirac or Weyl fermions with exotic properties, such as extremely high quantum mobility [1-7], large magnetoresistance [1-7] and potential topological superconductivity [8]. In Dirac (e.g., $Na_3Bi$ [9,10] and $Cd_3As_2$ [11-14]) and Weyl semimetals (e.g., transition metal monopnictides (Ta/Nb)(As/P) [15-22] and $(W/Mo)Te_2$ [23-32]), the linear energy bands cross at scattered points in momentum space, forming Dirac or Weyl nodes protected by crystal symmetry. In addition to these topological semimetals with discrete Dirac/Weyl nodes, another type of topological semimetal - the topological nodal line semimetal featuring Dirac bands crossing along a one-dimensional line/loop, has also been predicted [33-39] and experimentally observed in several material systems such as $(Pb/Tl)TaSe_2$ [40,41], *WHM* (*W*=Zr, Hf; *H*=Si, Ge, Sb; *M*=S, Se, Te) [42-50] and $PtSn_4$ [51].

Recently, there has been growing interests in relativistic fermions hosted by two-dimensional (2D) square or nearly square net in layered materials, such as $AMn(Bi/Sb)_2$ (A=Ca, Sr, Ba, or rare earth element) [52-66] and *WHM*-type [42-48] materials. In these materials, the 2D/quasi-2D relativistic fermions are harbored by the 2D square nets formed by the group VI or V elements including Bi, Sb, Si, and Ge, and exhibit distinct properties such as half-integer quantum Hall effect in $EuMnBi_2$ [59], interlayer quantum tunneling from the zeroth Landau level in $YbMnBi_2$ [64], 2D nonsymmorphic Dirac state in Zr(Si/Ge)*M* (*M*=S, Se, Te) [42,45], and tunable Weyl and Dirac states in CeSbTe [67]. These results imply that the layered topological semimetals provide a unique platform to explore novel topological fermion physics.

In addition to Bi, Sb, Si, and Ge layers mentioned above, a Sn square net is also expected to harbor relativistic fermions, as has been revealed in ZrSnTe [27]. As a member of the *WHM* material family, ZrSnTe also possesses tetragonal layered structure, with the Sn square plane sandwiched by the Zr-Te layers, forming the Te-Zr-Sn-Zr-Te slabs [Fig. 1(a)]. ZrSnTe has been predicted to be a weak topological insulator [68], or a nodal-line semimetal in the bulk form [38,45] but a 2D topological insulator in the monolayer form [38]. Signatures of topological bands on the top layer of bulk ZrSnTe has been probed by ARPES measurements [69]. However, the topological nature of bulk bands is yet to be experimentally clarified. In this work, we report de Haas–van Alphen (dHvA) quantum oscillation studies on ZrSnTe single crystals and show the features consistent with the predicted bulk topological fermions.

Unlike other *WHM* materials whose single crystals were usually synthesized using a chemical vapor transport method [42,44,45,47,49,70-74], single crystals of ZrSnTe [Fig. 1(b), inset] can be grown only using a flux method [69,70]. The ZrSnTe single crystals used in this work were grown using Sn flux. The starting materials of Zr, Sn, and Te with the molar ratio of 1:10:1 were sealed into a quartz tube under high vacuum. The reagents were heated to 1000 °C, kept at this temperature for 24 hours, and slowly cooled down to 400 °C. Plate-like ZrSnTe single crystals can be obtained after removing the excess Sn flux by centrifugation, as shown in the inset of Fig. 1(b). The compositions of the synthesized crystals were analyzed using energy dispersive spectroscopy (EDS). The excellent crystallinity is demonstrated by the sharp x-ray diffraction peaks as shown in Fig. 1(b), which can be indexed as the (00L) reflections according to the tetragonal structure with space group of P4/*nmm* for ZrSnTe. The lattice constant *c*, *i.e.* the slab thickness, is derived to be 8.721 Å, consistent with the previous reports [70]. In *WHM* compounds for which the square net is formed by the group 14 elements *H* (*H*=Si, Ge, Sn) [38] [Fig. 1(a)], increasing the ionic radius of *H* from Si to Ge and to Sn lengthens the *H-H* bonding distances and subsequently elongates the lattice constant *a* (*a* = *b*), with an accompanied increase of Te-Te distance [70]. This further reduces of the steric crowding of Te and allows Te to move toward Zr in the neighboring slab, resulting in enhanced interlayer coupling and 3D electronic structure for ZrSnTe [70]. Therefore, the interlay binding energy for ZrSnTe is the highest among *WHM* compounds with *W*=Zr/Hf, *H*=Si/Ge/Sn and *M*=S/Se/Te [38].

Signatures of topological Dirac states in ZrSnTe, including light effective mass, high mobility, and nontrivial Berry phase, have been revealed in our dHvA quantum oscillation studies using the 31T resistive magnet in NHMFL, Tallahassee. We have observed clear oscillations in magnetic torque measurements performed with a piezoresistive cantilever. Given the torque signal is expected to vanish when magnetic field is perfectly aligned along the out-of-plane (*B* // *c*) and in-plane (*B* // *ab*) directions, we performed the measurements with the fields nearly along the out-of-plane and in-plane directions (denoted by *c'* and *ab'* respectively). In Figs. 2(a) and 2(b), we have presented the field dependences of magnetic torque at different temperatures for ZrSnTe for *B*//*c'* and *B*//*ab'*, respectively. For both field orientations, strong dHvA oscillations occur for $B > 10$T at $T = 1.8$K, and remain visible until the temperature is increased to $T = 20$K. The presence of quantum oscillations for both field orientations indicates a 3D

Fermi surface in ZrSnTe despite its layered crystal structure, in agreement with the strong interlayer binding [38,70] of this material. Signature of Zeeman splitting, which has been probed for some *WHM* compounds including ZrSiS [73,75], ZrGeS [47], and ZrGeSe [47], is not clearly observed in ZrSnTe up to 31T for both field orientations.

From the oscillatory torque $\tau_{osc}$ for (c) $B//c'$ and (d) $B//ab'$ obtained by subtracting the smooth background [Figs. 2(c) and 2(d)], one can find oscillation patterns for both $B//c'$ and $B//ab'$ contain multiple frequency components. This can be clearly seen in the fast Fourier transform (FFT) analyses as shown in the insets of Figs. 2(c) and 2(d). For both field orientations, three major frequencies can be resolved: $F_\alpha = 28T$, $F_\beta = 238T$, $F_\gamma = 350T$ for $B//c'$; and $F_{\alpha'} = 17$ T, $F_\varepsilon = 78$ T, $F_\eta = 125$ T for $B//ab'$. Although there has been no report on the calculated Fermi surface for ZrSnTe, the first principle calculations [38,45] have revealed that the energy bands crossing near the Fermi level in ZrSnTe are all Dirac bands. Therefore, the probed oscillation frequencies should arise from topological non-trivial bands. The large oscillation frequency up to a few hundred teslas is rarely seen for Dirac and Weyl semimetals due to their point-like Fermi surface, but is a generic feature of *WHM* materials [44,47,71-73,75-78] due to the fact the nodal line is enclosed by a large Fermi surface.

We can obtain further information about the nodal-line fermions in ZrSnTe from the analyses of dHvA oscillations. The dHvA oscillations for a 3D system can be described by the 3D Lifshitz-Kosevich (LK) formula [79,80] with a Berry phase being taken into account for a topological system [81]:

$$\Delta\tau \propto -B^{1/2} R_T R_D R_S \sin[2\pi(\frac{F}{B}+\gamma-\delta)] \qquad (1)$$

where $R_T = \alpha T\mu/B\sinh(\alpha T\mu/B)$, $R_D = \exp(-\alpha T_D\mu/B)$, and $R_S = \cos(\pi g\mu/2)$. $\mu = m^*/m_0$ is the ratio of effective cyclotron mass $m^*$ to free electron mass $m_0$. $T_D$ is Dingle temperature, and $\alpha = (2\pi^2 k_B m_0)/(\hbar e)$. The oscillation is described by the sine term with a phase factor $\gamma - \delta$, in which $\gamma = \frac{1}{2} - \frac{\phi_B}{2\pi}$ and $\phi_B$ is Berry phase. The phase shift $\delta$ is determined by the dimensionality of the Fermi surface and has a value of $\pm 1/8$ for 3D cases, with the sign depending on whether the probed extreme cross-section area of the FS is maximal (-) or minimal (+) [79]. From the LK formula, the effective mass $m^*$ can be obtained through the fit of the temperature dependence of the oscillation amplitude by the thermal damping factor $R_T$. In the case of multi-frequency oscillations, the oscillation amplitude for each frequency can be represented by the amplitude of FFT peak (FFTA), and the parameter $1/B$ in $R_T$ should be the average inverse field $1/\bar{B}$, defined as $1/\bar{B} = (1/B_{max} + 1/B_{min})/2$, where $B_{max}$ and $B_{min}$ define the magnetic field range used for FFT. As shown in the insets of Figs. 3a and 3b, for all probed oscillation frequencies, the obtained effective masses are in the range of 0.16 - 0.21 $m_0$ (see Table 1), which are slightly larger than or comparable with those of other *WHM* materials obtained from dHvA oscillations [44,47,72,73] and agrees well with the nature of massless relativistic fermions.

High quantum mobility and π Berry phase are also important characteristics of topological fermions. For the multi-frequency oscillations in ZnSnTe, these parameters cannot be directly obtained through the conventional approaches, *i.e.*, the Dingle plot and the Landau Level fan diagram, but can be extracted through the fit of the oscillation pattern to the generalized multiband LK formula [82]. This method has been shown to be efficient for the analyses of the multi-frequency quantum oscillations in many *WHM* compounds [44,47,73]. For dHvA oscillations under nearly out-of-plane field ($B//c'$), the LK-fit was not very successful with all three major frequency components being included. Nevertheless, a better fitting result was obtained when the lower frequency ($F_\alpha$) component is filtered out, as shown in Fig. 3(c). With the effective mass and frequency as the known parameters, the multiband LK model reproduces the oscillation pattern well, yielding Dingle temperatures $T_D$ of 20.7 K and 46.8 K for the $\beta$ and $\gamma$ bands, respectively (Table I). The quantum relaxation time $\tau_q$ [$= \hbar/(2\pi k_B T_D)$] corresponding to these values of Dingle temperature are $5.9 \times 10^{-14}$ and $2.6 \times 10^{-14}$ s, from which the quantum mobility $\mu_q$ ($= e\tau/m^*$) are estimated to be 741 and 241 cm$^2$V$^{-1}$s$^{-1}$, respectively, for the $\beta$ and $\gamma$ bands (Table I). The obtained quantum mobility values for ZrSnTe are remarkably smaller than those of ZrSi*M* [44,73] and HfSiS [72] obtained from dHvA oscillations, but close to those of the ZrGe*M* compounds [47]. Given that high mobility is the generic feature of topological relativistic fermions [1], the low quantum mobility seen in ZrSnTe might be associated with the greater spin-orbit coupling (SOC)-induced gap, as will be discussed later.

The relativistic nature of carriers is generally manifested by the non-trivial Berry phases in quantum oscillations. The multiband LK-fit provides an effective approach for the determination of Berry phase, which is particularly useful for the multi-frequency oscillations [44,47,73,82]. From the LK-fit, we have obtained the phase factors ($\gamma-\delta$) for the $\beta$ and $\gamma$ bands, from which the Berry phase is determined to be (-0.62±0.25)$\pi$, and (-0.61±0.25)$\pi$, respectively (Table1). Such non-trivial Berry phase factors are consistent with the theoretically predicted topological Dirac states in all *WHM* systems [38].

Features of topological fermions are also extracted from the dHvA oscillations under the nearly in-plane field ($B//ab'$). For the three major bands α′, ε, and η, we have also obtained light effective masses of 0.17 $m_0$, 0.17 $m_0$, and 0.23 $m_0$, respectively, from the fits of the temperature dependence of FFT amplitude [Fig. 3(b)]. From the multiband LK-fits [Fig. 3(d)], we have also estimated the quantum mobility and non-trivial Berry phase for all three frequency components, as listed in Table I.

From the above analyses, we have revealed evidences of topological fermions in ZrSnTe which are consistent with the theoretically predicted topological nodal-line state in *WHM*-type compounds [38]. In *WHM* compounds, their overall electronic structures are predicted to be similar [38,45], except for small discrepancies caused by the variations of SOC strength and structural dimensionality. As stated above, the interlay coupling in ZrSnTe is expected to be strong due to steric-electronic balance, which should result in strong 3D character. This has been confirmed by the 3D Fermi surface morphology revealed in our angular dependence of the quantum oscillation measurements.

Using the measurement setup shown in the inset of Fig. 4b, we have collected dHvA oscillation data for ZrSnTe under different magnetic field orientations. The presence of dHvA oscillations in the whole angle range from $\theta=0°$ to $90°$ indicates 3D Fermi surface for ZrSnTe. After the background subtraction, the oscillation pattern displays a clear evolution with the rotation of the magnetic field, as shown in Fig. 4a. In Fig. 4c, we summarized the angular dependences of the major fundamental frequencies obtained from the FFT analyses. The continuous evolution of all probed oscillation frequencies in the whole angle range indicates that the observed quantum oscillations should be due to the bulk bands. The lowest frequency components probed under nearly out-of-plane and in-plane field (i.e., $F_\alpha$ and $F_{\alpha'}$) should arise from the same Fermi surface sheet, which appears to be less anisotropic given the weak angular dependence of the frequency. In contrast, the high frequencies display much stronger angular dependence. The difference of the probed two major high frequencies $F_\beta$ and $F_\gamma$ seen for $B//c'$ and the difference of $F_\varepsilon$ and $F_\eta$ seen for $B//ab'$ varies remarkably with angle, reaching zero at certain angles, which is suggestive of corrugated cylindrical Fermi surface sheets. The observed angular dependence of oscillation frequencies of ZrSnTe look similar to that of ZrGeS [47], implying that these two compounds possess similar 3D-like Fermi surface. Such a similarity in their band structures is consistent with the fact that they share similar crystallographic $c/a$ ratio (*i.e.*, 2.148 for ZrSnTe and 2.212 for ZrGeS) [70] and interlayer binding energy (i.e., 0.6861 eV for ZrSnTe and 0.6044 eV for ZrGeS) [38].

In addition to the structure dimensionality, SOC strength can also be fine-tuned in the *WHM* compounds with the different combinations of *W*, *H* and *M* elements. Unlike other ideal topological semimetals such as Na$_3$Bi [9,10], Cd$_3$As$_2$ [11-14], and (Ta/Nb)(As/P) [15-22] which possess symmetry-protected gapless Dirac and Weyl cones, the *C$_{2v}$* symmetry that leads to Dirac nodal-line in *WHM* allows for SOC gap [42], which results in massive Dirac fermions with reduced mobility. In ZrSiS, the exceptional properties including the very small effective mass and very high quantum mobility [44] is known to be associated with the tiny SOC gap (~20meV [42]) since element Si is light and gives rise to weak SOC. With increased SOC strength in other *WHM* compounds composed of heavier elements such as ZrSiSe/Te[44], ZrGe*M* [47], and HfSiS [72], heavier effective mass and reduced mobility have been observed in quantum oscillations. Given the larger total atomic number for ZrSnTe, it is not surprising to probe even heavier effective mass and much lower quantum mobility in our dHvA oscillation studies.

In conclusion, we have synthesized the single crystals of ZrSnTe and performed dHvA quantum oscillation studies on them. The analyses of dHvA quantum oscillation reveal properties consistent with the theoretically predicted bulk topological semimetal state for the first time, which further demonstrates that topological fermions can also be harbored by the Sn network. Owing to the strong interlayer coupling, the Dirac bands of ZrSnTe exhibit clear 3D nature despite its layered structure. The heavier effective mass and reduced quantum mobility seen in our analyses also sheds light on the effect of SOC on the Dirac bands in the *WHM*-type materials.

**ACKNOWLEDGMENTS**

J. H. is supported by the start-up fund from the University of Arkansas. Work at University of Tulane was supported by the US Department of Energy under grant DE-SC0014208

(support for personnel and material synthesis). The work at the National High Magnetic Field Laboratory is supported by the NSF Cooperative Agreement No. DMR-1157490 and the State of Florida (high field magnetic torque measurements).

**Reference**
[1] T. Liang, Q. Gibson, M. N. Ali, M. Liu, R. J. Cava, and N. P. Ong, *Ultrahigh mobility and giant magnetoresistance in the Dirac semimetal $Cd_3As_2$*, Nature Mater. **14**, 280 (2015).
[2] J. Xiong, S. Kushwaha, J. Krizan, T. Liang, R. J. Cava, and N. P. Ong, *Anomalous conductivity tensor in the Dirac semimetal $Na_3Bi$*, Europhys. Lett. **114**, 27002 (2016).
[3] C. Shekhar, A. K. Nayak, Y. Sun, M. Schmidt, M. Nicklas, I. Leermakers, U. Zeitler, Y. Skourski, J. Wosnitza, Z. Liu, Y. Chen, W. Schnelle, H. Borrmann, Y. Grin, C. Felser, and B. Yan, *Extremely large magnetoresistance and ultrahigh mobility in the topological Weyl semimetal candidate NbP*, Nature Phys. **11**, 645 (2015).
[4] N. J. Ghimire, L. Yongkang, M. Neupane, D. J. Williams, E. D. Bauer, and F. Ronning, *Magnetotransport of single crystalline NbAs*, J. Phys. Condens. Matter **27**, 152201 (2015).
[5] Z. Wang, Y. Zheng, Z. Shen, Y. Lu, H. Fang, F. Sheng, Y. Zhou, X. Yang, Y. Li, C. Feng, and Z.-A. Xu, *Helicity-protected ultrahigh mobility Weyl fermions in NbP*, Phys. Rev. B **93**, 121112 (2016).
[6] F. Arnold, C. Shekhar, S.-C. Wu, Y. Sun, R. D. dos Reis, N. Kumar, M. Naumann, M. O. Ajeesh, M. Schmidt, A. G. Grushin, J. H. Bardarson, M. Baenitz, D. Sokolov, H. Borrmann, M. Nicklas, C. Felser, E. Hassinger, and B. Yan, *Negative magnetoresistance without well-defined chirality in the Weyl semimetal TaP*, Nature Communications **7**, 11615 (2016).
[7] X. Huang, L. Zhao, Y. Long, P. Wang, D. Chen, Z. Yang, H. Liang, M. Xue, H. Weng, Z. Fang, X. Dai, and G. Chen, *Observation of the Chiral-Anomaly-Induced Negative Magnetoresistance in 3D Weyl Semimetal TaAs*, Phys. Rev. X **5**, 031023 (2015).
[8] H. Wang, H. Wang, H. Liu, H. Lu, W. Yang, S. Jia, X.-J. Liu, X. C. Xie, J. Wei, and J. Wang, *Observation of superconductivity induced by a point contact on 3D Dirac semimetal $Cd_3As_2$ crystals*, Nature Mater. **15**, 38 (2016).
[9] Z. Wang, Y. Sun, X.-Q. Chen, C. Franchini, G. Xu, H. Weng, X. Dai, and Z. Fang, *Dirac semimetal and topological phase transitions in $A_3Bi(A=Na, K, Rb)$*, Phys. Rev. B **85**, 195320 (2012).
[10] Z. K. Liu, B. Zhou, Y. Zhang, Z. J. Wang, H. M. Weng, D. Prabhakaran, S.-K. Mo, Z. X. Shen, Z. Fang, X. Dai, Z. Hussain, and Y. L. Chen, *Discovery of a Three-Dimensional Topological Dirac Semimetal, $Na_3Bi$*, Science **343**, 864 (2014).
[11] Z. Wang, H. Weng, Q. Wu, X. Dai, and Z. Fang, *Three-dimensional Dirac semimetal and quantum transport in $Cd_3As_2$*, Phys. Rev. B **88**, 125427 (2013).

[12]    S. Borisenko, Q. Gibson, D. Evtushinsky, V. Zabolotnyy, B. Büchner, and R. J. Cava, *Experimental Realization of a Three-Dimensional Dirac Semimetal*, Phys. Rev. Lett. **113**, 027603 (2014).

[13]    Z. K. Liu, J. Jiang, B. Zhou, Z. J. Wang, Y. Zhang, H. M. Weng, D. Prabhakaran, S. K. Mo, H. Peng, P. Dudin, T. Kim, M. Hoesch, Z. Fang, X. Dai, Z. X. Shen, D. L. Feng, Z. Hussain, and Y. L. Chen, *A stable three-dimensional topological Dirac semimetal $Cd_3As_2$*, Nature Mater. **13**, 677 (2014).

[14]    M. Neupane, S.-Y. Xu, R. Sankar, N. Alidoust, G. Bian, C. Liu, I. Belopolski, T.-R. Chang, H.-T. Jeng, H. Lin, A. Bansil, F. Chou, and M. Z. Hasan, *Observation of a three-dimensional topological Dirac semimetal phase in high-mobility $Cd_3As_2$*, Nature Commun. **5**, 3786 (2014).

[15]    S.-Y. Xu, I. Belopolski, N. Alidoust, M. Neupane, G. Bian, C. Zhang, R. Sankar, G. Chang, Z. Yuan, C.-C. Lee, S.-M. Huang, H. Zheng, J. Ma, D. S. Sanchez, B. Wang, A. Bansil, F. Chou, P. P. Shibayev, H. Lin, S. Jia, and M. Z. Hasan, *Discovery of a Weyl Fermion semimetal and topological Fermi arcs*, Science **349**, 613 (2015).

[16]    B. Q. Lv, H. M. Weng, B. B. Fu, X. P. Wang, H. Miao, J. Ma, P. Richard, X. C. Huang, L. X. Zhao, G. F. Chen, Z. Fang, X. Dai, T. Qian, and H. Ding, *Experimental Discovery of Weyl Semimetal TaAs*, Phys. Rev. X **5**, 031013 (2015).

[17]    B. Q. Lv, N. Xu, H. M. Weng, J. Z. Ma, P. Richard, X. C. Huang, L. X. Zhao, G. F. Chen, C. E. Matt, F. Bisti, V. N. Strocov, J. Mesot, Z. Fang, X. Dai, T. Qian, M. Shi, and H. Ding, *Observation of Weyl nodes in TaAs*, Nature Physics **11**, 724 (2015).

[18]    L. X. Yang, Z. K. Liu, Y. Sun, H. Peng, H. F. Yang, T. Zhang, B. Zhou, Y. Zhang, Y. F. Guo, M. Rahn, D. Prabhakaran, Z. Hussain, S. K. Mo, C. Felser, B. Yan, and Y. L. Chen, *Weyl semimetal phase in the non-centrosymmetric compound TaAs*, Nature Phys. **11**, 728 (2015).

[19]    S.-Y. Xu, N. Alidoust, I. Belopolski, Z. Yuan, G. Bian, T.-R. Chang, H. Zheng, V. N. Strocov, D. S. Sanchez, G. Chang, C. Zhang, D. Mou, Y. Wu, L. Huang, C.-C. Lee, S.-M. Huang, B. Wang, A. Bansil, H.-T. Jeng, T. Neupert, A. Kaminski, H. Lin, S. Jia, and M. Zahid Hasan, *Discovery of a Weyl fermion state with Fermi arcs in niobium arsenide*, Nature Phys. **11**, 748 (2015).

[20]    N. Xu, H. M. Weng, B. Q. Lv, C. E. Matt, J. Park, F. Bisti, V. N. Strocov, D. Gawryluk, E. Pomjakushina, K. Conder, N. C. Plumb, M. Radovic, G. Autes, O. V. Yazyev, Z. Fang, X. Dai, T. Qian, J. Mesot, H. Ding, and M. Shi, *Observation of Weyl nodes and Fermi arcs in tantalum phosphide*, Nature Commun. **7**, 11006 (2015).

[21]    H. Weng, C. Fang, Z. Fang, B. A. Bernevig, and X. Dai, *Weyl Semimetal Phase in Noncentrosymmetric Transition-Metal Monophosphides*, Phys. Rev. X **5**, 011029 (2015).

[22]    S.-M. Huang, S.-Y. Xu, I. Belopolski, C.-C. Lee, G. Chang, B. Wang, N. Alidoust, G. Bian, M. Neupane, C. Zhang, S. Jia, A. Bansil, H. Lin, and M. Z. Hasan, *A Weyl Fermion semimetal with surface Fermi arcs in the transition metal monopnictide TaAs class*, Nature Commun. **6**, 7373 (2015).

[23]    A. A. Soluyanov, D. Gresch, Z. Wang, Q. Wu, M. Troyer, X. Dai, and B. A. Bernevig, *Type-II Weyl semimetals*, Nature **527**, 495 (2015).

[24]    Y. Sun, S.-C. Wu, M. N. Ali, C. Felser, and B. Yan, *Prediction of Weyl semimetal in orthorhombic $MoTe_2$*, Phys. Rev. B **92**, 161107 (2015).

[25]    F. Y. Bruno, A. Tamai, Q. S. Wu, I. Cucchi, C. Barreteau, A. de la Torre, S. McKeown Walker, S. Riccò, Z. Wang, T. K. Kim, M. Hoesch, M. Shi, N. C. Plumb, E.

Giannini, A. A. Soluyanov, and F. Baumberger, *Observation of large topologically trivial Fermi arcs in the candidate type-II Weyl $WTe_2$*, Phys. Rev. B **94**, 121112 (2016).

[26] C. Wang, Y. Zhang, J. Huang, S. Nie, G. Liu, A. Liang, Y. Zhang, B. Shen, J. Liu, C. Hu, Y. Ding, D. Liu, Y. Hu, S. He, L. Zhao, L. Yu, J. Hu, J. Wei, Z. Mao, Y. Shi, X. Jia, F. Zhang, S. Zhang, F. Yang, Z. Wang, Q. Peng, H. Weng, X. Dai, Z. Fang, Z. Xu, C. Chen, and X. J. Zhou, *Observation of Fermi arc and its connection with bulk states in the candidate type-II Weyl semimetal $WTe_2$*, Phys. Rev. B **94**, 241119 (2016).

[27] Y. Wu, D. Mou, N. H. Jo, K. Sun, L. Huang, S. L. Bud'ko, P. C. Canfield, and A. Kaminski, *Observation of Fermi Arcs in Type-II Weyl Semimetal Candidate $WTe_2$*, Phys. Rev. B **94**, 121113(R) (2016).

[28] K. Deng, G. Wan, P. Deng, K. Zhang, S. Ding, E. Wang, M. Yan, H. Huang, H. Zhang, Z. Xu, J. Denlinger, A. Fedorov, H. Yang, W. Duan, H. Yao, Y. Wu, S. Fan, H. Zhang, X. Chen, and S. Zhou, *Experimental observation of topological Fermi arcs in type-II Weyl semimetal $MoTe_2$*, Nat. Phys. **12**, 1105 (2016).

[29] L. Huang, T. M. McCormick, M. Ochi, Z. Zhao, M.-T. Suzuki, R. Arita, Y. Wu, D. Mou, H. Cao, J. Yan, N. Trivedi, and A. Kaminski, *Spectroscopic evidence for a type II Weyl semimetallic state in $MoTe_2$*, Nat. Mater. **15**, 1155 (2016).

[30] J. Jiang, Z. K. Liu, Y. Sun, H. F. Yang, C. R. Rajamathi, Y. P. Qi, L. X. Yang, C. Chen, H. Peng, C. C. Hwang, S. Z. Sun, S. K. Mo, I. Vobornik, J. Fujii, S. S. P. Parkin, C. Felser, B. H. Yan, and Y. L. Chen, *Signature of type-II Weyl semimetal phase in $MoTe_2$*, Nature Communications **8**, 13973 (2017).

[31] A. Liang, J. Huang, S. Nie, Y. Ding, Q. Gao, C. Hu, S. He, Y. Zhang, C. Wang, B. Shen, J. Liu, P. Ai, L. Yu, X. Sun, W. Zhao, S. Lv, D. Liu, C. Li, Y. Zhang, Y. Hu, Y. Xu, L. Zhao, G. Liu, Z. Mao, X. Jia, F. Zhang, S. Zhang, F. Yang, Z. Wang, Q. Peng, H. Weng, X. Dai, Z. Fang, Z. Xu, C. Chen, and X. J. Zhou, *Electronic Evidence for Type II Weyl Semimetal State in $MoTe_2$*, arXiv:1604.01706 (2016).

[32] N. Xu, Z. J. Wang, A. P. Weber, A. Magrez, P. Bugnon, H. Berger, C. E. Matt, J. Z. Ma, and B. Q. L. B. B. Fu, N. C. Plumb, M. Radovic, E. Pomjakushina, K. Conder, T. Qian, J. H. Dil, J. Mesot, H. Ding, M. Shi, *Discovery of Weyl semimetal state violating Lorentz invariance in $MoTe_2$*, arXiv:1604.02116 (2016).

[33] Y. Kim, B. J. Wieder, C. L. Kane, and A. M. Rappe, *Dirac Line Nodes in Inversion-Symmetric Crystals*, Phys. Rev. Lett. **115**, 036806 (2015).

[34] H. Weng, Y. Liang, Q. Xu, R. Yu, Z. Fang, X. Dai, and Y. Kawazoe, *Topological node-line semimetal in three-dimensional graphene networks*, Phys. Rev. B **92**, 045108 (2015).

[35] L. S. Xie, L. M. Schoop, E. M. Seibel, Q. D. Gibson, W. Xie, and R. J. Cava, *A new form of $Ca_3P_2$ with a ring of Dirac nodes*, APL Mater. **3**, 083602 (2015).

[36] R. Yu, H. Weng, Z. Fang, X. Dai, and X. Hu, *Topological Node-Line Semimetal and Dirac Semimetal State in Antiperovskite $Cu_3PdN$*, Phys. Rev. Lett. **115**, 036807 (2015).

[37] M. Zeng, C. Fang, G. Chang, Y.-A. Chen, T. Hsieh, A. Bansil, H. Lin, and L. Fu, *Topological semimetals and topological insulators in rare earth monopnictides*, arXiv:1504.03492 (2015).

[38] Q. Xu, Z. Song, S. Nie, H. Weng, Z. Fang, and X. Dai, *Two-dimensional oxide topological insulator with iron-pnictide superconductor LiFeAs structure*, Phys. Rev. B **92**, 205310 (2015).

[39]     H. Huang, S. Zhou, and W. Duan, *Type-II Dirac fermions in the PtSe$_2$ class of transition metal dichalcogenides*, Phys. Rev. B **94**, 121117 (2016).
[40]     G. Bian, T.-R. Chang, R. Sankar, S.-Y. Xu, H. Zheng, T. Neupert, C.-K. Chiu, S.-M. Huang, G. Chang, I. Belopolski, D. S. Sanchez, M. Neupane, N. Alidoust, C. Liu, B. Wang, C.-C. Lee, H.-T. Jeng, C. Zhang, Z. Yuan, S. Jia, A. Bansil, F. Chou, H. Lin, and M. Z. Hasan, *Topological nodal-line fermions in spin-orbit metal PbTaSe$_2$*, Nature Commun. **7**, 10556 (2016).
[41]     G. Bian, T.-R. Chang, H. Zheng, S. Velury, S.-Y. Xu, T. Neupert, C.-K. Chiu, S.-M. Huang, D. S. Sanchez, I. Belopolski, N. Alidoust, P.-J. Chen, G. Chang, A. Bansil, H.-T. Jeng, H. Lin, and M. Z. Hasan, *Drumhead Surface States and Topological Nodal-Line Fermions in TlTaSe$_2$*, Phys. Rev. B **93**, 121113 (2016).
[42]     L. M. Schoop, M. N. Ali, C. Straszer, A. Topp, A. Varykhalov, D. Marchenko, V. Duppel, S. S. P. Parkin, B. V. Lotsch, and C. R. Ast, *Dirac cone protected by non-symmorphic symmetry and three-dimensional Dirac line node in ZrSiS*, Nature Commun. **7**, 11696 (2016).
[43]     M. Neupane, I. Belopolski, M. M. Hosen, D. S. Sanchez, R. Sankar, M. Szlawska, S.-Y. Xu, K. Dimitri, N. Dhakal, P. Maldonado, P. M. Oppeneer, D. Kaczorowski, F. Chou, M. Z. Hasan, and T. Durakiewicz, *Observation of Topological Nodal Fermion Semimetal Phase in ZrSiS*, Phys. Rev. B **93**, 201104 (2016).
[44]     J. Hu, Z. Tang, J. Liu, X. Liu, Y. Zhu, D. Graf, K. Myhro, S. Tran, C. N. Lau, J. Wei, and Z. Mao, *Evidence of Topological Nodal-Line Fermions in ZrSiSe and ZrSiTe*, Phys. Rev. Lett. **117**, 016602 (2016).
[45]     A. Topp, J. M. Lippmann, A. Varykhalov, V. Duppel, B. V. Lotsch, C. R. Ast, and L. M. Schoop, *Non-symmorphic band degeneracy at the Fermi level in ZrSiTe*, New J. Phys. **18**, 125014 (2016).
[46]     M. M. Hosen, K. Dimitri, I. Belopolski, P. Maldonado, R. Sankar, N. Dhakal, G. Dhakal, T. Cole, P. M. Oppeneer, D. Kaczorowski, F. Chou, M. Z. Hasan, T. Durakiewicz, and M. Neupane, *Tunability of the topological nodal-line semimetal phase in ZrSiX-type materials (X=S, Se, Te)*, Phys. Rev. B **95**, 161101 (2017).
[47]     J. Hu, Y. L. Zhu, D. Graf, Z. J. Tang, J. Y. Liu, and Z. Q. Mao, *Quantum oscillation studies of topological semimetal candidate ZrGeM (M = S, Se, Te)*, Phys. Rev. B **95**, 205134 (2017).
[48]     D. Takane, Z. Wang, S. Souma, K. Nakayama, C. X. Trang, T. Sato, T. Takahashi, and Y. Ando, *Dirac-node arc in the topological line-node semimetal HfSiS*, Phys. Rev. B **94**, 121108 (2016).
[49]     C. Chen, X. Xu, J. Jiang, S. C. Wu, Y. P. Qi, L. X. Yang, M. X. Wang, Y. Sun, N. B. M. Schröter, H. F. Yang, L. M. Schoop, Y. Y. Lv, J. Zhou, Y. B. Chen, S. H. Yao, M. H. Lu, Y. F. Chen, C. Felser, B. H. Yan, Z. K. Liu, and Y. L. Chen, *Dirac line nodes and effect of spin-orbit coupling in the nonsymmorphic critical semimetals MSiS (M=Hf,Zr)*, Phys. Rev. B **95**, 125126 (2017).
[50]     M. M. Hosen, G. Dhakal, K. Dimitri, P. Maldonado, A. Aperis, F. Kabir, P. M. Oppeneer, D. Kaczorowski, T. Durakiewicz, and M. Neupane, *Observation of topological nodal-line fermionic phase in GdSbTe*, arXiv:1707.05292 (2017).
[51]     Y. Wu, L.-L. Wang, E. Mun, D. D. Johnson, D. Mou, L. Huang, Y. Lee, S. L. Bud'ko, P. C. Canfield, and A. Kaminski, *Dirac node arcs in PtSn$_4$*, Nature Phys. **12**, 667 (2016).

[52]    J. Park, G. Lee, F. Wolff-Fabris, Y. Y. Koh, M. J. Eom, Y. K. Kim, M. A. Farhan, Y. J. Jo, C. Kim, J. H. Shim, and J. S. Kim, *Anisotropic Dirac Fermions in a Bi Square Net of SrMnBi$_2$*, Phys. Rev. Lett. **107**, 126402 (2011).

[53]    Y. Feng, Z. Wang, C. Chen, Y. Shi, Z. Xie, H. Yi, A. Liang, S. He, J. He, Y. Peng, X. Liu, Y. Liu, L. Zhao, G. Liu, X. Dong, J. Zhang, C. Chen, Z. Xu, X. Dai, Z. Fang, and X. J. Zhou, *Strong Anisotropy of Dirac Cones in SrMnBi$_2$ and CaMnBi$_2$ Revealed by Angle-Resolved Photoemission Spectroscopy*, Sci. Rep. **4** (2014).

[54]    K. Wang, D. Graf, H. Lei, S. W. Tozer, and C. Petrovic, *Quantum transport of two-dimensional Dirac fermions in SrMnBi$_2$*, Phys. Rev. B **84**, 220401 (2011).

[55]    K. Wang, D. Graf, L. Wang, H. Lei, S. W. Tozer, and C. Petrovic, *Two-dimensional Dirac fermions and quantum magnetoresistance in CaMnBi$_2$*, Phys. Rev. B **85**, 041101 (2012).

[56]    K. Wang, D. Graf, and C. Petrovic, *Quasi-two-dimensional Dirac fermions and quantum magnetoresistance in LaAgBi$_2$*, Phys. Rev. B **87**, 235101 (2013).

[57]    W. Yi-Yan, Y. Qiao-He, and X. Tian-Long, *Large linear magnetoresistance in a new Dirac material BaMnBi$_2$*, Chinese Physics B **25**, 107503 (2016).

[58]    L. Li, K. Wang, D. Graf, L. Wang, A. Wang, and C. Petrovic, *Electron-hole asymmetry, Dirac fermions, and quantum magnetoresistance in BaMnBi$_2$*, Phys. Rev. B **93**, 115141 (2016).

[59]    H. Masuda, H. Sakai, M. Tokunaga, Y. Yamasaki, A. Miyake, J. Shiogai, S. Nakamura, S. Awaji, A. Tsukazaki, H. Nakao, Y. Murakami, T.-h. Arima, Y. Tokura, and S. Ishiwata, *Quantum Hall effect in a bulk antiferromagnet EuMnBi$_2$ with magnetically confined two-dimensional Dirac fermions*, Sci. Adv. **2**, e1501117 (2016).

[60]    M. A. Farhan, L. Geunsik, and S. Ji Hoon, *AEMnSb$_2$ (AE=Sr, Ba): a new class of Dirac materials*, J. Phys. Condens. Matter **26**, 042201 (2014).

[61]    J. Y. Liu, J. Hu, Q. Zhang, D. Graf, H. B. Cao, S. M. A. Radmanesh, D. J. Adams, Y. L. Zhu, G. F. Cheng, X. Liu, W. A. Phelan, J. Wei, M. Jaime, F. Balakirev, D. A. Tennant, J. F. DiTusa, I. Chiorescu, L. Spinu, and Z. Q. Mao, *A magnetic topological semimetal Sr$_{1-y}$Mn$_{1-z}$Sb$_2$ (y, z < 0.10)*, Nature Mater. **16**, 905 (2017).

[62]    J. Liu, J. Hu, H. Cao, Y. Zhu, A. Chuang, D. Graf, D. J. Adams, S. M. A. Radmanesh, L. Spinu, I. Chiorescu, and Z. Mao, *Nearly Massless Dirac fermions hosted by Sb square net in BaMnSb$_2$*, Sci. Rep. **6**, 30525 (2016).

[63]    S. Borisenko, D. Evtushinsky, Q. Gibson, A. Yaresko, T. Kim, M. N. Ali, B. Buechner, M. Hoesch, and R. J. Cava, *Time-Reversal Symmetry Breaking Type-II Weyl State in YbMnBi$_2$*, arXiv:1507.04847 (2015).

[64]    J. Y. Liu, J. Hu, D. Graf, T. Zou, M. Zhu, Y. Shi, S. Che, S. M. A. Radmanesh, C. N. Lau, L. Spinu, H. B. Cao, X. Ke, and Z. Q. Mao, *Unusual interlayer quantum transport behavior caused by the zeroth Landau level in YbMnBi$_2$*, Nature Communications **8**, 646 (2017).

[65]    R. Kealhofer, S. Jang, S. M. Griffin, C. John, K. A. Benavides, S. Doyle, T. Helm, P. J. W. Moll, J. B. Neaton, J. Y. Chan, J. D. Denlinger, and J. G. Analytis, *Observation of two-dimensional Fermi surface and Dirac dispersion in YbMnSb$_2$*, arXiv:1708.03308 (2017).

[66]    J. B. He, Y. Fu, L. X. Zhao, H. Liang, D. Chen, Y. M. Leng, X. M. Wang, J. Li, S. Zhang, M. Q. Xue, C. H. Li, P. Zhang, Z. A. Ren, and G. F. Chen, *Quasi-two-dimensional massless Dirac fermions in CaMnSb$_2$*, Phys. Rev. B **95**, 045128 (2017).

[67]   L. M. Schoop, A. Topp, J. Lippmann, F. Orlandi, L. Muechler, M. G. Vergniory, Y. Sun, A. W. Rost, V. Duppel, M. Krivenkov, S. Sheoran, P. Manuel, A. Varykhalov, B. Yan, R. K. Kremer, C. R. Ast, and B. V. Lotsch, *Tunable Weyl and Dirac states in the nonsymmorphic compound CeSbTe*, arXiv:1707.03408 (2017).
[68]   B. Bradlyn, L. Elcoro, J. Cano, M. G. Vergniory, Z. Wang, C. Felser, M. I. Aroyo, and B. A. Bernevig, *Topological quantum chemistry*, Nature **547**, 298 (2017).
[69]   R. Lou, J. Z. Ma, Q. N. Xu, B. B. Fu, L. Y. Kong, Y. G. Shi, P. Richard, H. M. Weng, Z. Fang, S. S. Sun, Q. Wang, H. C. Lei, T. Qian, H. Ding, and S. C. Wang, *Emergence of topological bands on the surface of ZrSnTe crystal*, Phys. Rev. B **93**, 241104 (2016).
[70]   C. Wang and T. Hughbanks, *Main Group Element Size and Substitution Effects on the Structural Dimensionality of Zirconium Tellurides of the ZrSiS Type*, Inorg. Chem. **34**, 5524 (1995).
[71]   R. Singha, A. Pariari, B. Satpati, and P. Mandal, *Large nonsaturating magnetoresistance and signature of nondegenerate Dirac nodes in ZrSiS*, Proc. Natl. Acad. Sci. USA **114**, 2468 (2017).
[72]   N. Kumar, K. Manna, Y. Qi, S.-C. Wu, L. Wang, B. Yan, C. Felser, and C. Shekhar, *Unusual magnetotransport from Si-square nets in topological semimetal HfSiS*, Phys. Rev. B **95**, 121109(R) (2017).
[73]   J. Hu, Z. Tang, J. Liu, Y. Zhu, J. Wei, and Z. Mao, *Nearly massless Dirac fermions and strong Zeeman splitting in the nodal-line semimetal ZrSiS probed by de Haas-van Alphen quantum oscillations*, Phys. Rev. B **96**, 045127 (2017).
[74]   R. Sankar, G. Peramaiyan, I. P. Muthuselvam, C. J. Butler, K. Dimitri, M. Neupane, G. N. Rao, M. T. Lin, and F. C. Chou, *Crystal growth of Dirac semimetal ZrSiS with high magnetoresistance and mobility*, Sci. Rep. **7**, 40603 (2017).
[75]   M. Matusiak, J. R. Cooper, and D. Kaczorowski, *Thermoelectric quantum oscillations in ZrSiS*, Nature Communications **8**, 15219 (2017).
[76]   M. N. Ali, L. M. Schoop, C. Garg, J. M. Lippmann, E. Lara, B. Lotsch, and S. Parkin, *Butterfly Magnetoresistance, Quasi-2D Dirac Fermi Surfaces, and a Topological Phase Transition in ZrSiS*, Sci. Adv. **2**, e1601742 (2016).
[77]   S. Pezzini, M. R. v. Delft, L. Schoop, B. Lotsch, A. Carrington, M. I. Katsnelson, N. E. Hussey, and S. Wiedmann, *Unconventional mass enhancement around the Dirac nodal loop in ZrSiS*, arXiv:1701.09119 (2017).
[78]   X. Wang, X. Pan, M. Gao, J. Yu, J. Jiang, J. Zhang, H. Zuo, M. Zhang, Z. Wei, W. Niu, Z. Xia, X. Wan, Y. Chen, F. Song, Y. Xu, B. Wang, G. Wang, and R. Zhang, *Evidence of Both Surface and Bulk Dirac Bands and Anisotropic Nonsaturating Magnetoresistance in ZrSiS*, Advanced Electronic Materials **2**, 1600228 (2016).
[79]   I. M. Lifshitz and A. M. Kosevich, *Theory of Magnetic Susceptibility in Metals at Low Temperatures*, Sov. Phys. JETP **2**, 636 (1956).
[80]   D. Shoenberg, *Magnetic Oscillations in Metals* (Cambridge Univ. Press, Cambridge, 1984).
[81]   G. P. Mikitik and Y. V. Sharlai, *Manifestation of Berry's Phase in Metal Physics*, Phys. Rev. Lett. **82**, 2147 (1999).
[82]   J. Hu, J. Y. Liu, D. Graf, S. M. A. Radmanesh, D. J. Adams, A. Chuang, Y. Wang, I. Chiorescu, J. Wei, L. Spinu, and Z. Q. Mao, *π Berry phase and Zeeman splitting of Weyl semimetal TaP*, Sci. Rep. **6**, 18674 (2016).

**Table I**. Parameters derived from the analyses of dHvA oscillations for ZrSnTe. $F$, oscillation frequency; $T_D$, Dingle temperature; $m^*$, effective mass; $\tau$, quantum relaxation time; $\mu_q$, quantum mobility; $\phi_B$, Berry phase.

|  | $F$ (T) | $m^*/m_0$ | $T_D$ (K) | $\tau$ (ps) | $\mu$ (cm$^2$V$^{-1}$s$^{-1}$) | $\phi_B$ |
|---|---|---|---|---|---|---|
| | 28 | 0.18 | - | - | - | - |
| $B//c'$ | 238 | 0.16 | 20.7 | 0.059 | 741 | $(0.62\pm0.25)\pi$ |
| | 350 | 0.21 | 46.8 | 0.026 | 241 | $(0.61\pm0.25)\pi$ |
| | 17 | 0.17 | 15.2 | 0.080 | 828 | $(-0.49\pm0.25)\pi$ |
| $B//ab'$ | 78 | 0.17 | 25.1 | 0.048 | 496 | $(-0.48\pm0.25)\pi$ |
| | 125 | 0.23 | 21.3 | 0.057 | 436 | $(0.70\pm0.25)\pi$ |

# Figure captions

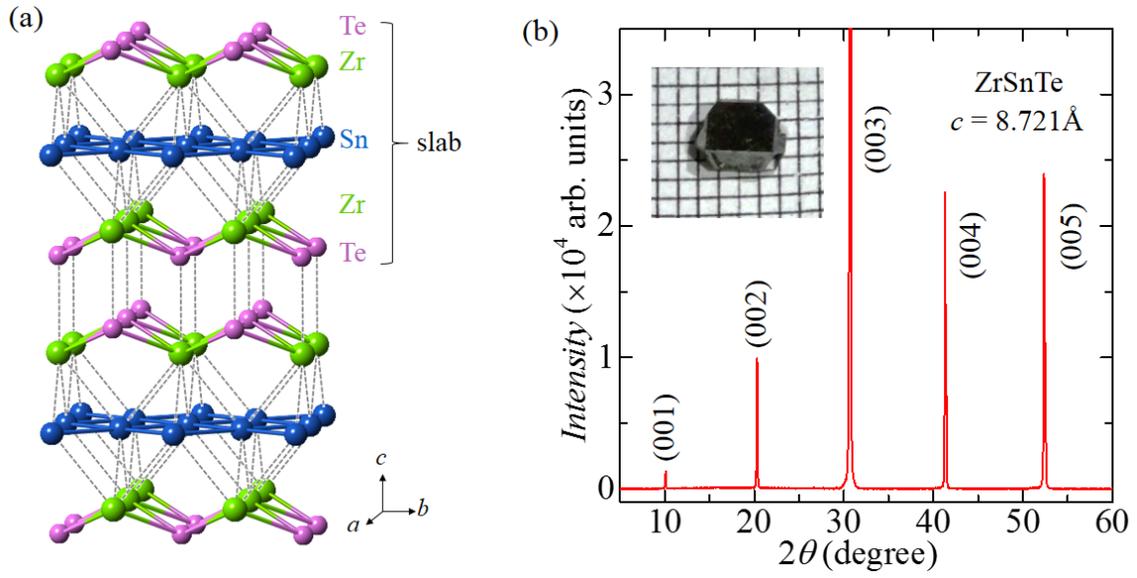

FIG. 1. (a) Crystal structure of ZrSnTe. (b) Single crystal x-ray diffraction patterns for ZrSnTe. Inset: an image of a ZrSnTe single crystal.

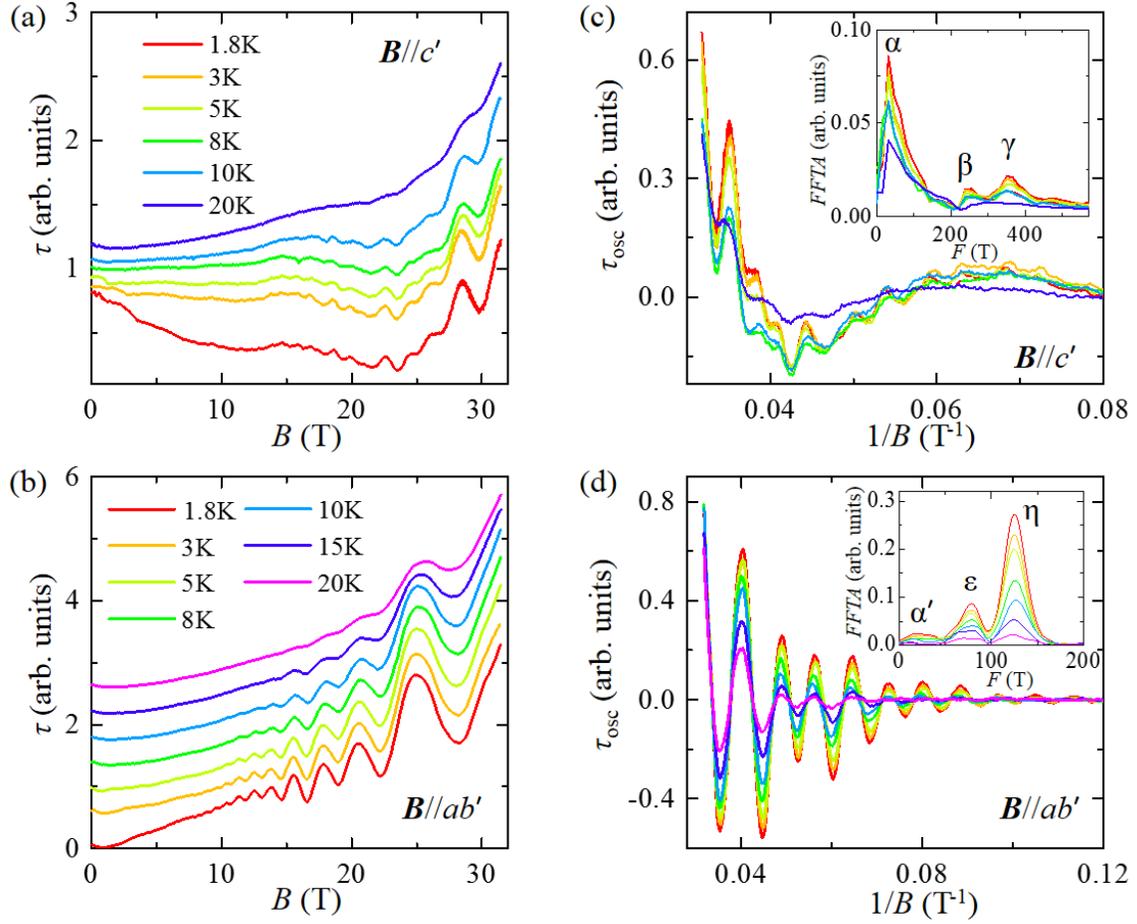

FIG. 2. (a) and (b): The field dependence of magnetic torque $\tau$ for ZrSnTe at different temperatures, for the magnetic field along the (a) nearly out-of-plane direction ($B//c'$) and (b) nearly in-plane direction ($B//ab'$). (c) and (d): The oscillatory component $\tau_{osc}$ for (c) $B//c'$ and (d) $B//ab'$. Inset: FFT for $\tau_{osc}$ for (c) $B//c'$ and (d) $B//ab'$.

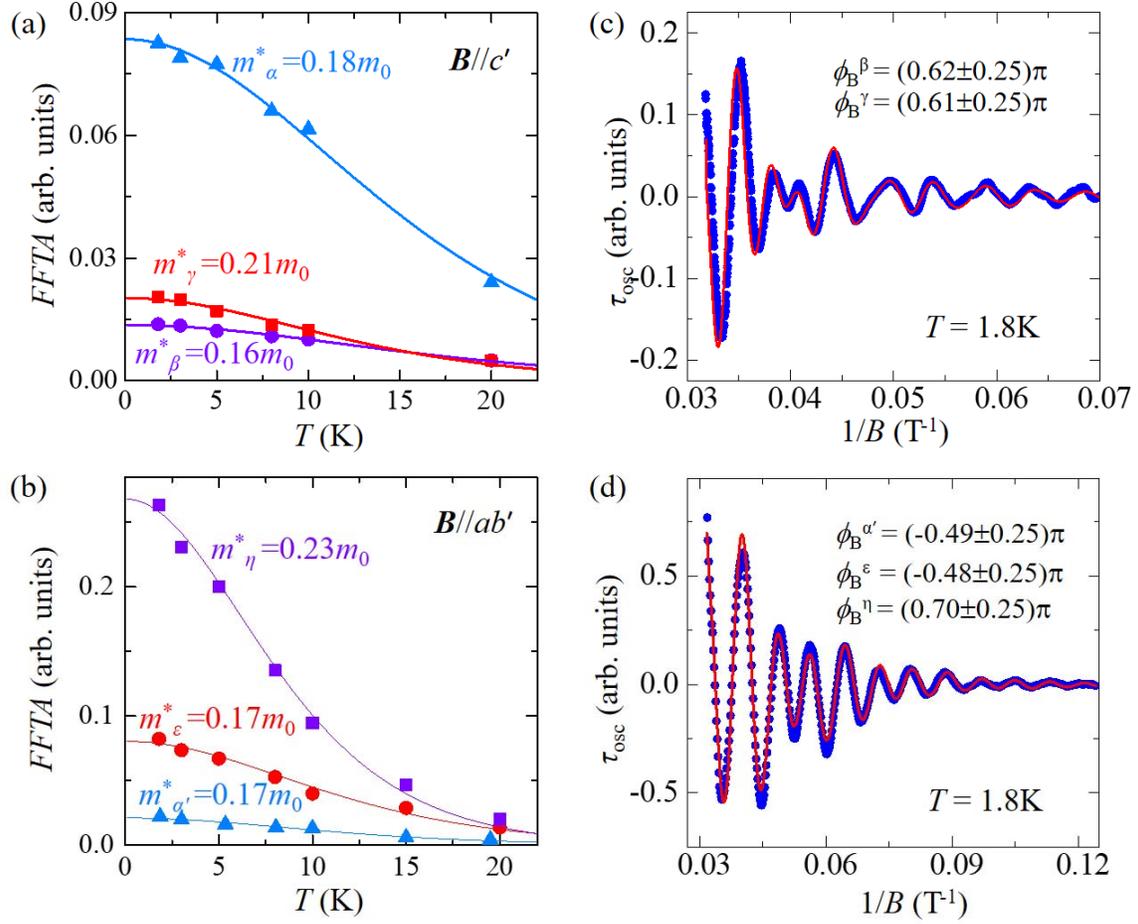

FIG. 3. (a) and (b) The temperature dependence of the FFT amplitude (FFTA) of the major fundamental frequencies for (a) $B//c'$ and (b) $B//ab'$. The fits to the LK formula (solid lines) yield effective mass. (c) The fit of the oscillation pattern for $B//c'$ at $T=1.8$K to the multi-band LK formula. The lower-frequency ($F_\alpha$) component has been filtered out. (d) The fit of the oscillation pattern for $B// ab'$ at $T=1.8$K to the multi-band LK formula.

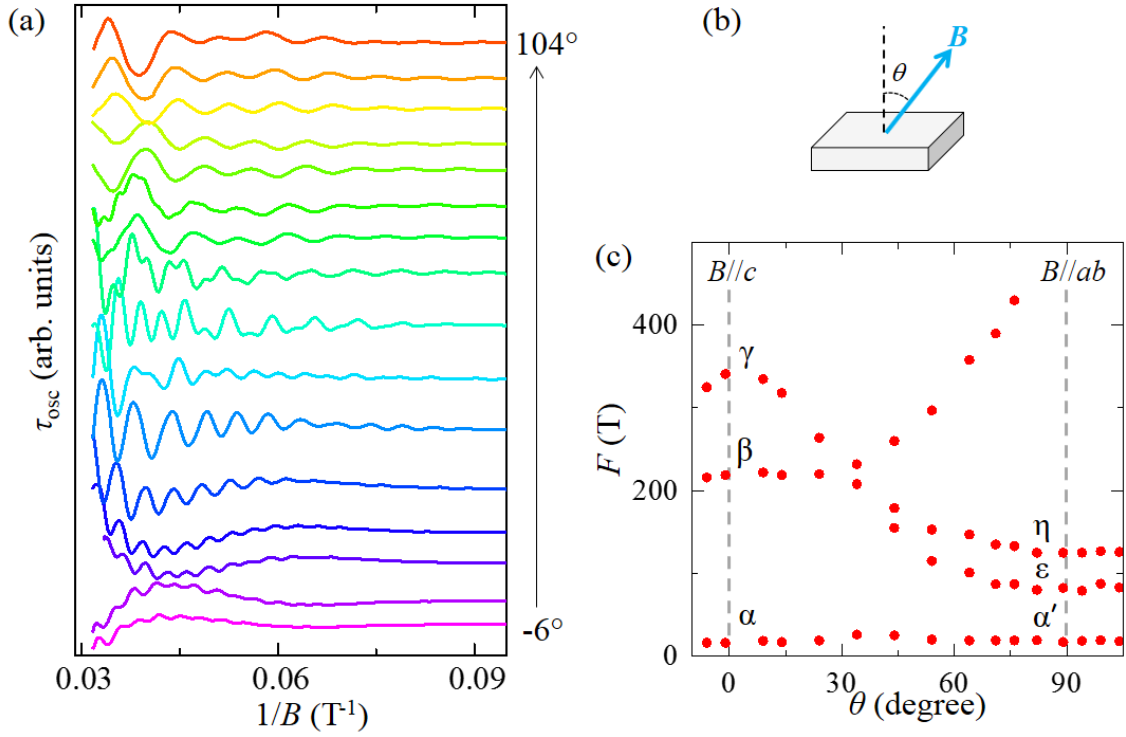

FIG. 4. (a) dHvA oscillations of ZrSnTe at $T$=1.8K under different magnetic field orientations. (b) Schematic of measurement setup. (c) The angular dependence of oscillation frequencies for ZrSnTe.